\documentclass[12pt]{article}
\usepackage{epsfig}

\parskip=\medskipamount 
plus.3em minus.5em \abovedisplayshortskip=.5em plus.2em minus.4em
\renewcommand{\thesection}{\arabic{section}}

\catcode`@=11

\@addtoreset{equation}{section} \@addtoreset{equation}{subsection}
\def\theequation{\ifnum\value{section}=0 \arabic{equation}\ignorespaces
\else \ifnum\value{section}=-1 A.\arabic{equation}\ignorespaces
\else \ifnum\value{subsection}=0
\thesection.\arabic{equation}\ignorespaces \else
\thesection.\arabic{subsection}.\arabic{equation}\ignorespaces
                             \fi
                        \fi
                   \fi}

{\catcode`\'=\active \def'{{}^\bgroup\prim@s}}

\catcode`@=12

\newcommand{\bq}{\begin{equation}}
\newcommand{\be}{\begin{equation}}
\newcommand{\fq}{\end{equation}}
\newcommand{\ee}{\end{equation}}
\newcommand{\bqr}{\begin{eqnarray}}
\newcommand{\beqs}{\begin{eqnarray}}
\newcommand{\fqr}{\end{eqnarray}}
\newcommand{\eeqs}{\end{eqnarray}}

\newcommand{\rf}[1]{(\ref{#1})}

\def\bop#1{\setbox0=\hbox{$#1M$}\mkern1.5mu
    \vbox{\hrule height0pt depth.04\ht0
    \hbox{\vrule width.04\ht0 height.9\ht0 \kern.9\ht0
    \vrule width.04\ht0}\hrule height.04\ht0}\mkern1.5mu}
\def\Box{{\mathpalette\bop{}}}                        

\begin{document}
\thispagestyle{empty}

\begin{flushright}
\begin{tabular}{l}
hep-th/0503141 \\
\end{tabular}
\end{flushright}

\vskip .6in
\begin{center}

{\bf Comment on the Riemann Hypothesis}

\vskip .6in

{\bf Gordon Chalmers}
\\[5mm]

{e-mail: gordon@quartz.shango.com}

\vskip .5in minus .2in

{\bf Abstract}

\end{center}

The Riemann hypothesis is identified with zeros of ${\cal N}=4$ supersymmetric gauge 
theory four-point amplitude.  The zeros of the $\zeta(s)$ function are identified with 
th complex dimension of the spacetime, or the dimension of the toroidal compactification.  
A sequence of dimensions are identified in order to map the zeros of the amplitude to 
the Riemann hypothesis.

\vfill\break

\section{Introduction} 

The century old Riemann hypothesis \cite{WaWh} states that the only nontrivial zeros of the 
zeta function, 

\bqr  
\zeta(s) = \sum_{n=1}^\infty {1\over n^s} = \prod (1-p^{-s})^{-1} \ , 
\fqr 
are on the set of points $s={1\over 2}+it$.  Tremendous numerical computations support 
this conjecture.  The purpose of this article is to identify that under certain conditions 
imposed on the ${\cal N}=4$ amplitude, the zeros of the Riemann zeta function are found 
in a formal sense with the zeros of these amplitudes.  
In precise terms, after an identification of the real parts of a sequence of derived 
dimensions, all gauge theory amplitudes vanish when the zeta function has zeros on  
the real axis $s=1/2 + it$.  (The Riemann zeta function on this axis has some similarities 
with the vanishing of the partition function of certain condensed matter theories as a 
function of couplings, i.e. Lee-Yang zeros.) 

\section{Review of the S-duality derivative expansion} 

The ${\cal N}=4$ spontaneously broken theory is examined in this work.  The Lagrangian is, 

\bqr  
{\cal S}={1\over g^2} {\rm Tr}~ \int ~ d^4x \bigl[ F^2 + \phi \Box \phi + 
  \psi {\slash D} \psi + \left[ \phi,\phi\right]^2 \bigr] \ . 
\fqr 
The quantum theory is believed to have a full S-duality, which means that the gauge 
amplitudes are invariant: under $A\rightarrow A_D$ and $\tau\rightarrow (a\tau+b)/ 
(c\tau+d)$ the functional form of the amplitude is invariant.  The series supports a 
tower of dyonic muliplets satisfying the mass formula $m^2=2\vert n^i a_i +m^i a_{d,i}\vert^2$ 
with $a_i$ and $a_{d,i}$ the vacuum values of the scalars and their duals; $a_{d,i}=\tau 
a_i$.  The two couplings parameterizing the simplest SU(2)$\rightarrow$U(1) theory is, 

\bqr  
{\theta\over 2\pi} + {4\pi i\over g^2}=\tau=\tau_1+i\tau_2 \ , 
\fqr 
taking values in the Teichmuller space of the keyhole region in the upper half plane, 
i.e. $\vert\tau\vert\geq 1/2$ and $\vert \tau_1\vert \leq 1/2$.  The S-duality invariant 
scattering within the derivative expansion is constructed in \cite{Chalmers1}.  Derivative 
expansions in general are examined in \cite{Chalmers1}-\cite{Chalmers10}.  

The full amplitudes of ${\cal N}=4$ theory may be constructed either in a gauge coupling 
perturbative series, i.e. the usual diagrammatic expansion formulated via unitarity methods, 
or as an expansion in derivatives, with the latter approach being nonperturbative in 
coupling.  Both expansions are equivalent, found from a diagram by diagram basis.  

The full set of operators to create a spontaneously broken ${\cal N}=4$ gauge theory 
amplitude is found from 

\bqr 
{\cal O}= \prod_{j=1} {\rm Tr} F^k_j \ , 
\fqr 
with possible $\ln^{m_1}(\Box) \ldots \ln^{m_n})$ (from the massless sector) and 
combinations with the covariant derivative; the derivatives are gauge covariantized 
and the tensor contractions are implied.  The dimensionality of the operator is 
compensated by a factor of the 
vacuum expectation value, $\langle\phi^2\rangle^m$.  The generic tensor has been 
suppressed in the combination, and we did not include the fermions of scalars as in 
\cite{Chalmers1} because the gauge vertices are only required (the coefficients of 
course are found via the sewing, involving the integrations \cite{Chalmers1}, 
\cite{Chalmers3}-\cite{Chalmers5}).  

The generating function of the gauge theory ${\cal N}=4$ four-point amplitude is 
given 

\bqr 
{\cal S}_4 =\sum ~ \int d^dx~ h_n(\tau,\bar\tau) {\cal O}_n \ , 
\fqr 
with the ring of functions spanning $h_n(\tau,\bar\tau)$ consisting of the elements, 

\bqr 
\prod E_{s_j}^{(q_j,-q_j)} (\tau,\bar\tau) \ , 
\fqr 
and their weights  

\bqr 
\sum_j s_j = n/2 \ , \qquad \sum_j q_j = 0 \ , 
\fqr 
with $s=m/2+1$, and $n$ the number of gauge bosons.  The general covariant term in 
the effective theory has terms,  

\bqr 
\prod_{i=1}^{n_\partial} \nabla_{\mu_{\sigma'(i)}} \qquad  
\prod^{m_i^A} A_{\mu_{\sigma(i),a_{\sigma(i)}}} \prod_{j=1}^{n_i^\phi} 
 \phi_{a_{\rho(j)}} \prod^{m_i^\psi} \psi_{a_{\kappa(j)}}  \ , 
\fqr 
with the derivatives placed in various orderings (multiplying fields and products of 
combinations of fields; this is described in momentum space in \cite{Chalmers1}).  The  
multiplying Eisenstein series possessing weights, 

\bqr  
s=n_A+n_\phi+n_\psi/2 + n_\partial/2+2 \qquad q=n_\psi/2 \ . 
\fqr 
These terms span the general operator ${\cal O}$ in the generating functional.  The 
non-holomorphic weight $q$ is correlated with the R-symmetry.  

The perturbative coupling structure, for the gauge bosons as an example, has the 
form, 

\bqr 
g^{n-1} (g^2)^{n_{\rm max}/2} \Bigl[ \bigl({1\over g^2}\bigr)^{{\rm max}/2} , 
\ldots , \bigl({1\over g^2}\bigr)^{-n_{\rm max}/2+1} \Bigr] \ . 
\label{couplingexp}
\fqr 
The factor in brackets agrees with the modular expansion of the Eisenstein series 
pertinent to the scattering amplitudes, and the prefactor may be absorbed by a field 
redefinition, 

\bqr 
A\rightarrow g^{-2} A \qquad x\rightarrow g x \ , 
\fqr 
which maps the gauge field part of the Lagrangian into 

\bqr 
\int d^4x~ {1\over g^2} ~ {\rm Tr}\left( \partial A + {1\over g} A^2\right)^2 \ . 
\fqr 
This field redefinition, together with the supersymmetric completion, agrees with the 
${\cal N}=4$ S-duality self-mapping in a manifest way (the factor in front may be 
removed by a Weyl rescaling).  

Fermionic (and mixed) amplitudes would have a non-vanishing $q_j$ sum.  The Eisenstein 
functions have the representation 

\bqr 
E_{s_j}^{(q_j,-q_j)} (\tau,\bar\tau) = \sum_{(p,q)\neq (0,0)}  {\tau_2^s\over 
 (p+q\tau)^{s-q} (p+q\bar\tau)^{s+q}} \ , 
\fqr   
with an expansion containing two monomial terms and an infinite number of exponential 
(representing instanton) terms, 

\bqr 
E_s(\tau,\bar\tau) = 2\zeta(2s) \tau_2^s + {\sqrt\pi} {\Gamma(s-1/2)\over \Gamma(s)} 
\zeta(2s-1) \tau_2^{1-2s} + {\cal O}(e^{-2\pi\tau}) \ldots 
\fqr 
with a modification in the non-holomorphic counterpart, $E_s^{(q,-q)}$, but with the 
same zeta function factors.  The latter terms correspond to gauge theory instanton 
contributions to the amplitude; via S-duality all of the instantonic terms are available 
from the perturbative sector.  (At $s=0$ or $s={1\over 2}$ the expansion is finite: 
$\zeta(0)=-1$ and both $\zeta(2s-1)\vert_{s=0}$ and $\Gamma(s)\vert_{s=0}$ have simple 
poles.)  The $n$-point amplitudes, with the previously discussed modular weight, are 

\bqr 
\langle A(k_1) \ldots A(k_n)\rangle = \sum_q h_q^{(n)}(\tau,\bar\tau) f_q(k_1,\ldots,  
  k_n) \ , 
\fqr 
where the modular factor is h (with the weights $n_A/2+2$) and the kinematic structure 
of the higher derivative term $f_q$.  The $n_{\rm max}$ follows from the modular 
expansion $n_A/2+n_\partial/2+2$, and corresponds to a maximum loop contribution of 
$n_A+n_\partial+1$.  

We shall not review in detail the sewing relations that allow for a determination 
of the coefficients of the modular functions at the various derivative orders.  This 
is discussed in detail in \cite{Chalmers3}-\cite{Chalmers5}.  

\section{Rescaling of coupling} 

A rescaling of the coupling constant via $g\rightarrow g^{1+\epsilon}$ changes the 
expansion in \rf{couplingexp} to, 

\bqr 
(g^2)^{2+\epsilon} (g^2)^{(n_{\rm max}/2)(1+\epsilon)} \bigl[ 
 (g^2)^{(n_{\rm max}/2)(1+\epsilon)}, \ldots, (g^2)^{(-n_{\rm max}/2+1)(1+\epsilon)} 
\bigr] \ .  
\fqr 
The rescaling of the couplings into the metric and the gauge fields would naively 
generate a derivative expansion with modular functions labeled by $E_{s(1+\epsilon)}$, 
and hence different coefficients for the expansion.  These terms can always be 
supersymmetrized to obtain the remaining couplings involving the fermions and scalars.  
Within the loop expansion the zeta function takes values in accord with the dimension 
of the loop integrals, which suggests that the theory is in a different dimension from 
$4$ to $4(1+\epsilon/2)$; comparison with the loop expansion is required to determine 
this (note that the tree-level terms found from the first term in \rf{couplingexp} are 
invariant after including the gauge field rescaling; this is true for the scatteing 
after changing dimension).  

Note that for $\epsilon=-1$ the entire scattering has no coupling dependence; gauge 
theory in $d=2$ is topological, and the gauge field and coupling may be gauged away 
in a background without topology.  The self-consistency via the sewing knocks out 
the coefficients of the covariant gauge field operators and one is left with the 
scalar interactions; the fermionic terms vanish as they only couple to the gauge field.  
The dimension changes as $4(1+\epsilon/2)$, or rather to a dimension of $4(1+(d-4)/2)= 
4(-1+d/2)=-4+2d$.  

In the altered theory the ring of functions consists of 

\bqr 
\prod E_{s_i(1+\epsilon)}(\tau,\bar\tau) \qquad \sum s_i = s \ , 
\fqr 
with $s=n/2+1$, and $n$ being the number of external gauge bosons.  The expansion at 
$\epsilon=-1$ has finite coefficients.  

\section{Amplitudes and zeros of the Riemann function} 

The arguments of the Riemann zeta function for a given derivative term of the gauge 
theory scattering amplitude are $2s$ and $2s-1$.  In terms of $s=(n+2)/2$ the 
arguments of the zeta function are 

\bqr 
2(-2+d)(n+2) \qquad {\rm and} \qquad 2(-2+d)(n+2)-1   \ . 
\fqr 
If all of the real parts of the dimensions 

\bqr  
d_R = {1\over 4(n+2)}+2 \ , \qquad d_R = {3\over 4(n+2)} + 2 
\fqr 
are identified then the arguments of the zeta functions are on the real $s=1/2$ axis.  
These series have $d=2$ as a limit point, with a maximum dimension of $2+1/8=2.125$.  
The gauge sector vanishes for $d=2$, i.e. at the limit point.  

If the amplitudes vanished via the identification on the $s=1/m$ axis, then the real 
part of the dimension would be 

\bqr 
d_R={1\over 2m(n+2)}+2 \ , \qquad d_R={3\over 2m(n+2)} + 2 \ . 
\label{dimensions}
\fqr 
Example dimensions pertaining to the Riemann hypothesis, $m=2$ in \rf{dimensions}, are 

\bqr  
d_R = 2+1/12=25/12, \qquad 2+1/16=33/16, \qquad 2+1/20=41/20 \ , 
\fqr 
\bqr 
d_R = 2+1/4=9/4 , \qquad 2+3/16=35/16 , 2+3/20=43/20 \ . 
\fqr 
The identification can be thought of as toroidal compactification with the dimensions 
identified, or as a series of identified four-manifolds.  

\section{Discussion} 

${\cal N}=4$ supersymmetric gauge theory amplitudes, including the nonperturbative 
corrections, are examined as a function of complex dimension.  The zeros of the Riemann 
zeta function enforce the vanishing of the four-point gauge theory amplitudes.  More 
precisely, the Riemann hypothesis is equivalent to the vanishing of the amplitudes 
of ${\cal N}=4$ four-point functions when the theory is dimensionally reduced on 
identified tori of dimension $d$, with $d=id_I+d_R$, 

\bqr 
d_R={1\over 2m(n+2)}+2 \qquad {\rm and} \qquad d_R={3\over 2m(n+2)}+2 \ . 
\fqr 
The real parts of these dimensions range from $2$ to $2.125$, with $d=2$ ($d_I=0$) special 
from the point of the triviality of the gauge field (pure gauge).

\vfill\break

\end{document}